**Boron isotope effect on the thermal conductivity of boron arsenide single crystals**

Haoran Sun[1], Ke Chen[2], Geethal Amila Gamage[1], Hamidreza Ziyaee[3], Fei Wang[1], Yu Wang[1,4], Viktor G. Hadjiev[3], Fei Tian[1]†, Gang Chen[2], and Zhifeng Ren[1]†

[1] *Department of Physics and Texas Center for Superconductivity (TcSUH), University of Houston, Houston, Texas 77204, USA*

[2] *Department of Mechanical Engineering, Massachusetts Institute of Technology, Cambridge, MA 02139, USA*

[3] *Department of Mechanical Engineering and Texas Center for Superconductivity (TcSUH), University of Houston, Houston, Texas 77204, USA*

[4] *Institute of Advanced Materials, Hubei Normal University, Huangshi, Hubei 435002, China*

†To whom correspondence should be addressed, email: feifei131305@163.com, zren@uh.edu

**Abstract**

Boron arsenide (BAs) with a zinc blende structure has recently been discovered to exhibit unusual and ultrahigh thermal conductivity ($\kappa$), providing a new outlook for research on BAs and other high thermal conductivity materials. Technology for BAs crystal growth has been continuously improving, however, the influence of boron isotopes, pure or mixed, on the thermal conductivity in BAs is still not completely clear. Here we report detailed studies on the growth of single crystals of BAs with different isotopic ratios and demonstrate that the $\kappa$ of isotopically pure BAs is at least 10% higher than that of BAs grown from natural B. Raman spectroscopy characterization shows differences in scattering among various BAs samples. The presented results will be helpful in guiding further studies on the influence of isotopes on optimizing $\kappa$ in BAs.

**Introduction**

The heat produced in high power density electronic devices imposes a major limitation on the performance of these devices. Cooling of the devices can be done by heat transfer in three ways: conduction, convection, and radiation [1]. The most efficient way for heat transferring away of electronic devices is by using materials with high thermal conductivity ($\kappa$). Normally, heat is conducted by electrons ($\kappa_e$) and phonons ($\kappa_l$) [2]. The room-temperature (RT) record for materials with high isotropic $\kappa_e$ is held by copper (Cu, 483 W m$^{-1}$ K$^{-1}$) [3] and the highest RT isotropic $\kappa_l$ is measured in diamond (2290 W m$^{-1}$ K$^{-1}$) [4,5]. However, the cost of diamond is still high due to difficulties in producing crystals with large enough size, whereas Cu's thermal conductivity is not high enough for many applications and its high electrical conductivity usually interferes with the circuits in electronic devices. Thus, finding suitable materials with ultrahigh isotropic $\kappa$ for heat dissipation in high power density electronic devices is imperative.

Since 2013, boron arsenide (BAs) has been at the forefront of research on ultrahigh isotropic $\kappa$ materials. BAs research was originally initiated by the results of first-principles calculations [6-8] that predicted BAs thermal conductivity comparable to that of diamond. Early experimental efforts showed BAs samples with relatively low but promising $\kappa$ values [9,10] and the later developments delivered BAs with $\kappa$ values [11-13] as high as the theoretically predicted $\kappa$ ~1300 W m$^{-1}$ K$^{-1}$ at RT when four-phonon scattering is taken into consideration [14,15]. BAs $\kappa$ value is smaller than that of natural diamond (2290 W m$^{-1}$ K$^{-}$ [1]) [4], but still much higher than that of any other reported bulk material. In addition, BAs is a semiconductor with an indirect band gap of about 1.8 eV at RT [16-18]. Such bulk zinc blende semiconductors with high isotropic thermal conductivity, unlike the two-dimensional materials, *e.g.*, PbTe, with low across-plane $\kappa$ [19], are highly welcomed for use in electronic designs. Currently, BAs is the only known bulk semiconductor exhibiting isotropic thermal conductivity that exceeds 1000 W m$^{-1}$ K$^{-1}$. At the same time, the mechanical performance of BAs [20] and its coefficient of thermal expansion (4.1x10$^{-6}$ K$^{-1}$ at RT) [21] are comparable to most of the widely used semiconductors, such as GaAs, Si, and SiC, indicating the great potential of BAs as a promising future material for the electronics industry.

Experimental achievements reported to date have generally been focused on BAs synthesized by using the naturally occurring boron isotope mixture $^{nat}$B (19.9% $^{10}$B and 80.1% $^{11}$B) [11-13]. As a key factor that affects the lattice vibration, and thus the $\kappa$, isotopes have been found to have a significant influence on the $\kappa$ of high $\kappa$ materials such as diamond and cubic boron nitride [22,23]. For BAs, the use of different pure isotopes is supposed to have lesser effect on $\kappa$ values than BN [24]. Indeed, in the kinetic theory of thermal conductivity, $\kappa$ is a sum over all phonon modes of a product of the phonon specific heat, the phonon transport lifetime that depends on the phonon-phonon, phonon-isotope, and electron-phonon scattering rates, and the square of the phonon (group) velocities. The acoustic phonons have the highest velocities and the isotope effect will be seen stronger if they involve vibrations of isotopes. Since the atomic mass of As is much higher than that of B, As atoms dominate the heat-carrying acoustic phonons. Varying B isotopes in BAs changes the acoustic phonon dispersion and phonon velocities negligibly little. The two stable B-isotopes, $^{10}$B and $^{11}$B, affect only the optical phonons. Under the three-phonon scattering mechanism, the RT $\kappa$ of isotopically pure BAs ($^{pure}$BAs) is more than 3100 W m$^{-1}$ K$^{-1}$ [6,7], nearly 50% higher than that of $^{nat}$BAs under the same assumption. However, when four-phonon scattering processes are taken into account in the refined theoretical predictions [14], the RT $\kappa$ value of $^{pure}$BAs is only 20% higher than that of $^{nat}$BAs, departing significantly from the conclusion of the three-phonon processes [24]. Therefore, understanding the isotope effects on the $\kappa$ value of BAs would help to confirm the correctness of the four-phonon processes in the prediction of $\kappa$ by first-principles calculations. Since $^{75}$As is the only known stable As isotope, we focus here on the effect of boron isotopes.

In this paper, we report the isotope effects in BAs experimentally investigated on $^{nat}$BAs, $^{10}$BAs, and $^{11}$BAs single crystals grown by the chemical vapor transport (CVT) method [13]. The $\kappa$ values of different BAs samples were compared and Raman spectra of phonons involving isotopes were tested and analyzed. We expect these results will help to clarify the influence of isotopes on the thermal properties of BAs single crystals.

**Results and discussion**

$^{10}$B (boron powder, ≥96% atomic $^{10}$B, crystalline, particle size: from 0.425 mm to 4.75 mm, 99.9% metals basis, Alfa Aesar), $^{11}$B (boron powder, ≥96% atomic $^{11}$B, crystalline, particle size: less than 0.15 mm, 99.9% metals basis, Alfa Aesar), and $^{nat}$B (boron powder, crystalline, particle size: from 0.425 mm to 4.75 mm, 99.9999% metals basis, Alfa Aesar) were used as different boron sources. Manually mixed B powders of 20% $^{10}$B and 80% $^{11}$B (denoted $^{10}$B$_{0.2}$$^{11}$B$_{0.8}$), and 50% $^{10}$B and 50% $^{11}$B (denoted $^{10}$B$_{0.5}$$^{11}$B$_{0.5}$) were also prepared as boron sources for comparison with results obtained from natural B. Details of the CVT growth process can be found in previous reports [25,26]. $^{10}$BAs and $^{11}$BAs single crystals with dimensions of 2.0 x 1.0 x 0.1 mm$^3$ were obtained within a two-week growth time, as shown in Figure 1A and 1B, respectively. The transparency and dark reddish color of the single crystals clearly show that their RT band gap is around 1.78 eV, which matches recent calculations [16-18]. Their smaller size compared to the reported best $^{nat}$BAs grown by the same method is likely due to the differences of the $^{10}$B and $^{11}$B sources from the natural B source used for the larger single crystals reported previously [13].

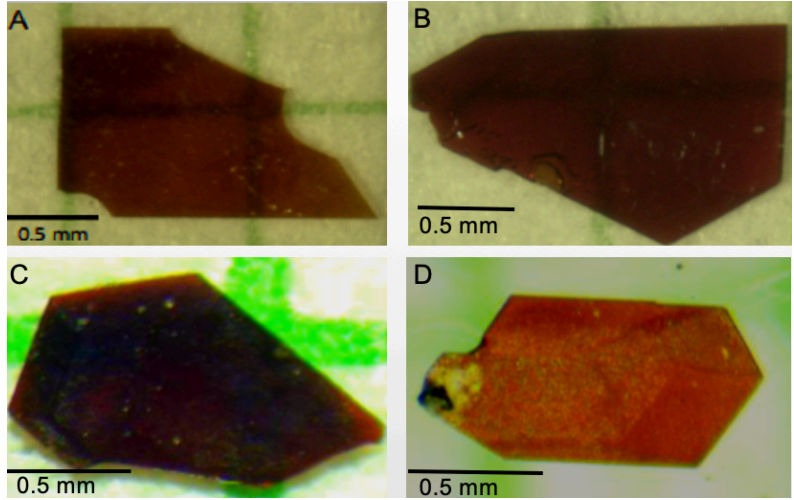

**Figure 1**. Optical microscopy images of (A) $^{10}$BAs, (B) $^{11}$BAs, (C) $^{10}$B$_{0.2}$$^{11}$B$_{0.8}$As, and (D) $^{10}$B$_{0.5}$$^{11}$B$_{0.5}$As single crystals.

The time domain thermoreflectance (TDTR) method [27-29] was used to measure the $\kappa$ of the grown $^{10}$BAs and $^{11}$BAs samples. TDTR has been validated to effectively measure a material's thermal conductivities, especially for mm- or µm-scale samples, where traditional heat transport measurements cannot be applied due to the near-zero temperature difference over a short length. A set of TDTR measurements was taken following the previously reported procedure [13], where the diameters (1/e$^2$) of the pump and probe laser spots are 58 µm and 9 µm, respectively. Pump intensity was modulated by an electro optic modulator (EOM) at 3-MHz frequency. An 82 nm -thick aluminum coating was deposited to the surface of each BAs crystal. $\kappa$ values for each sample are measured on many spots and the highest and lowest values together with the calculated values are summarized in Table 1. The highest RT $\kappa$ obtained from $^{10}$BAs (Figure 1A) is 1260 ± 130 W m$^{-1}$ K$^{-1}$, higher than that of the $^{nat}$BAs sample (1160 ± 130 W m$^{-1}$ K$^{-1}$) measured by the same method [13], which shows that isotopically pure $^{10}$BAs exhibits slightly higher $\kappa$ as compared to $^{nat}$BAs and the measured $\kappa$ is in agreement with the predicted value assuming 4-phonon scattering. The highest RT $\kappa$ of $^{11}$BAs (1180 ± 130 W m$^{-1}$ K$^{-1}$) (Figure 1B) is smaller than that of $^{10}$BAs. For comparison, we also grew BAs single crystals by hand-mixing B isotopes in different ratios. Unfortunately, the highest $\kappa$ values obtained from the as-prepared $^{10}$B$_{0.2}$$^{11}$B$_{0.8}$As (Figure 1C) and $^{10}$B$_{0.5}$$^{11}$B$_{0.5}$As (Figure 1D) crystals are less than 300 W m$^{-1}$ K$^{-1}$, much lower than expected, which is probably due to the extra step of manual mixing of different B isotopes and some unknown reasons, further studies are in progress. Each spot was measured at least 30 times to guarantee accuracy.

**Table 1**. Thermal conductivity of isotopically pure BAs.

| Composition | Label in Figure 1 | Highest $\kappa$ (W m$^{-1}$ K$^{-1}$) | Lowest $\kappa$ (W m$^{-1}$ K$^{-1}$) | Calculated $\kappa$ (W m$^{-1}$ K$^{-1}$) [24] |
|---|---|---|---|---|
| $^{10}$BAs | Fig. 1A | 1260 | 689 | 1374 |
| $^{11}$BAs | Fig. 1B | 1180 | 381 | 1331 |
| $^{nat}$BAs [13] | | 1160 | 640 | 1188 |

We used high-resolution Raman spectroscopy to study the phonon dynamics of BAs samples with boron isotope content. The Raman spectra of BAs samples were measured using a Horiba JY T64000 triple spectrometer excited by a 632 nm laser at 294 K [30]. The spectra were collected in a back-scattering configuration with incident and scattered light directions normal to the large face of the crystals in Figure 1. The Raman spectra of all BAs samples shown in Figure 1 are displayed in Figure 2. BAs has only one phonon that is active in the first-order Raman scattering and seen as a single Raman peak because of its very small LO-TO splitting [30]. In Figure 2B the experimental Raman peak for $^{10}$BAs is at 726 cm$^{-1}$, close to the calculated value of 729 cm$^{-1}$, and that of $^{11}$BAs is at 698 cm$^{-1}$, which is also in good agreement with the calculated value of 699 cm$^{-1}$ [30]. In the mixed $^{10}$B$_{(1-x)}$$^{11}$B$_x$As (x = 0.8 or 0.5) samples, the frequency of this phonon changes as $\omega_x$ is approximately equal to $[(1-x) \cdot m(^{10}B) + x \cdot m(^{11}B)]^{-0.5}$, where $m(^{10}B)$ and $m(^{11}B)$ are the masses of the respective boron isotopes and x is the portion of $^{11}$B in the isotope mixture. This expression provides a good approximation to the positions of the sharp peaks in Figure 2B. In boron mixed isotope samples with prevailing $^{11}$B content, there also exists isotope disorder that relaxes the Raman selection rules, such that $^{10}$B vibrations from other points of the Brillouin zone become Raman active and produce Raman bands in addition to the main Raman peak. These vibrations originate in the area of overlapping of $^{10}$BAs and $^{11}$BAs phonon density of states (phDOS) [30]. The overlapping of frequencies and wavevectors of $^{10}$B and $^{11}$B vibrations is maximized in the $^{10}$B$_{0.5}$$^{11}$B$_{0.5}$As sample, which in turn produces a broad Raman band (two shoulders) around the main peak with frequency $\sim [0.5 \cdot m(^{10}B) + 0.5 \cdot m(^{11}B)]^{-0.5}$. New modelling approaches, however, are needed to reveal completely the origin of this band.

The Raman scattering from phonons is a two-photon process in which incoming laser light is exciting electrons in material, which in turn results in emission (Stokes) or absorption (Anti-Stokes) of phonons via electron-phonon coupling. Laser excitations can also produce Raman scattering from electronic excitations, known as electronic Raman scattering (ERS). It has been most notable in doped semiconductors as p-doped Si. Electronic Raman scattering has been observed also in a number of BAs samples. ERS produces a scattering background and Fano lineshape of the Raman peaks in BAs [11]. The appearance of ERS correlates with the hole concentration in BAs and it is found detrimental on the thermal conductivity [11]. We attribute this effect to the enhanced phonon scattering due to electron-phonon coupling. In BAs, the absence of ERS is notably manifested by appearance of several second-order Raman bands seen clearly in Figure 2(C) and (D). The assignment of the harmonics and combinational Raman bands in Figure 2 is straightforward given the phonon dispersion in Ref. [30].

For BAs samples a subtle Raman detail may provide additional information on the crystal quality. Even a slight structural disorder may relax the Raman scattering selection rules and produce a first-order Raman scattering band that resembles part or the whole phonon density of states. We present such an example in Figure 2C (dashed rectangle) in which the phDOS of acoustic modes around 200 cm$^{-1}$ becomes Raman active.

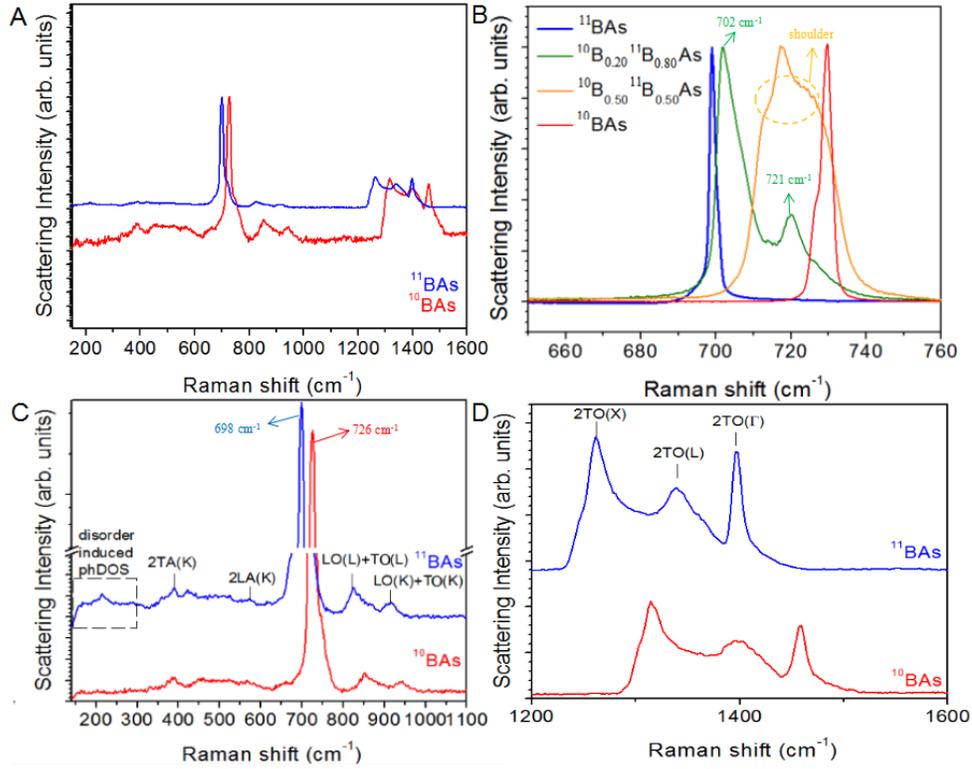

**Figure 2**. (A) Raman spectra for $^{10}$BAs and $^{11}$BAs samples over the entire frequency range. (B) Expanded view of Raman spectra in 650 – 760 cm$^{-1}$ for BAs samples with different boron-source ratios (pure $^{10}$BAs, $^{10}$B$_{0.2}$$^{11}$B$_{0.8}$As, $^{10}$B$_{0.5}$$^{11}$B$_{0.5}$As, and pure $^{11}$BAs). (C) Raman spectra for $^{10}$BAs and $^{11}$BAs samples at low frequency. (D) Raman spectra for $^{10}$BAs and $^{11}$BAs samples at high frequency.

In summary, the thermal conductivity of BAs single crystals grown with different boron isotopes was characterized by TDTR and the structures were analyzed by Raman spectroscopy. The BAs samples grown from the $^{10}$B source consistently exhibit higher thermal conductivity than those from $^{11}$B. Both pure isotope BAs have higher thermal conductivity than that of natural or mixed isotope B. Raman spectrometry displays obvious change in the phonon frequency of the isotopically pure BAs and indicates that B isotopes affect the phonon interactions. Further crystal quality improvement by eliminating the defects is needed to establish a more precise relationship between thermal conductivity and isotope content.

**Acknowledgment**

This work was supported by the Office of Naval Research under MURI grant N00014-16-1-2436.